\documentclass[showpacs,aps,twocolumn]{revtex4}

\usepackage{bm}
\usepackage{amsmath}
\usepackage{graphicx}
\usepackage{subfigure}
\usepackage[usenames,dvipsnames]{color}
\definecolor{darkblue}{RGB}{0,0,196}
\usepackage[colorlinks=true,linkcolor=darkblue,citecolor=darkblue,urlcolor=darkblue]{hyperref}

\usepackage{setspace}
\usepackage{hyperref}
\usepackage{footmisc}
\usepackage[makeroom]{cancel}
\usepackage{comment}
\usepackage{lineno}

\def\be{\begin{equation}}
\def\ee{\end{equation}}
\def\ba{\begin{eqnarray}}
\def\ea{\end{eqnarray}}

\usepackage{graphicx}
\usepackage{amsmath,bbm}
\usepackage{amssymb,bm}
\usepackage{yfonts}

\begin{document}

\title{Limiting fragmentation in high-energy nuclear collisions at the CERN Large Hadron Collider}

\author{Pragati~Sahoo}
\author{Pooja~Pareek}
\author{Swatantra~Kumar~Tiwari}
\author{Raghunath~Sahoo}
 \email{Raghunath.Sahoo@cern.ch}
\affiliation{Discipline of Physics, School of Basic Sciences, \\Indian Institute of Technology Indore, Simrol, Khandwa Road, Indore- 453552, INDIA}

\begin{abstract}
\noindent
The hypothesis of limiting fragmentation (LF) or it is called otherwise recently, as extended longitudinal scaling, is an interesting phenomena in high energy multiparticle production process. This paper discusses about different regions of phase space and their importance in hadron production, giving special emphasis on the fragmentation region. Although it was conjectured as a universal phenomenon in high energy physics, with the advent of higher center-of-mass energies, it has become prudent to analyse and understand the validity of such hypothesis in view of the increasing inelastic nucleon-nucleon cross-section ($\sigma_{\rm in}$). 
In this work, we revisit the phenomenon of limiting fragmentation for nucleus-nucleus (A+A) collisions in the pseudorapidity distribution of charged particles at various energies. We use energy dependent $\sigma_{\rm in}$ to transform the charged particle pseudorapidity distributions ($dN^{\rm AA}_{ch}/d\eta$) into differential cross-section per unit pseudorapidity ($d\sigma^{\rm AA}/d\eta$) of charged particles and study the phenomenon of LF. We find that in $d\sigma^{\rm AA}/d\eta$ LF seems to be violated at LHC energies while considering the energy dependent $\sigma_{\rm in}$. We also perform a similar study using A Multi-Phase Transport (AMPT) Model with string melting scenario and also find that LF is violated at LHC energies. 


\end{abstract}
\date{\today}
\maketitle 
\section{Introduction}
Understanding the particle productions in high energy nuclear collisions is always fascinating. The particle production in high energy collisions happens from three different regions: the projectile, the target and the central region. Particles emitted from the outer region are called projectile/target fragments. There are various nuclear fragmentation mechanisms discussed in literature~\cite{Jipa:2004fx, Kittel}. The most important are: a sudden fragmentation by explosive mechanisms, such as shock waves~\cite{Jipa:2004fx} and a slow fragmentation by the ``fission" of the spectator regions, mainly because of the interactions with the particles or fragments emitted from the participant region at transverse angles in the center-of-momentum system~\cite{Jipa:2004fx}. The latter is a purely low-energy nuclear phenomenon, where as the former is more applicable to relativistic domain of energies. 
During the late 1960s, the hypothesis of limiting fragmentation became important to understand the particle production~\cite{Kittel,Benecke:1969sh}. According to this hypothesis the produced particles, in the rest frame of one of the projectiles become independent of centre-of-mass energies, thus following a possible scaling (as a function of $\eta' = \eta \pm y_{\rm beam}$), known as limiting fragmentation (LF). As (pseudo)rapidity is a longitudinal variable it is also called longitudinal scaling. Here $y_{\rm beam} = \ln ({\sqrt{s_{\rm NN}}}/m_{\rm p})$, is beam rapidity and $m_{\rm p}$ is the mass of proton. There have been several attempts to understand the nature of hadronic interactions which lead to limiting fragmentation and the deviations from it~\cite{Gelis:2006tb, JalilianMarian:2002wq, Brogueira:2006nz}. 

\begin{figure}
\includegraphics[height=22em]{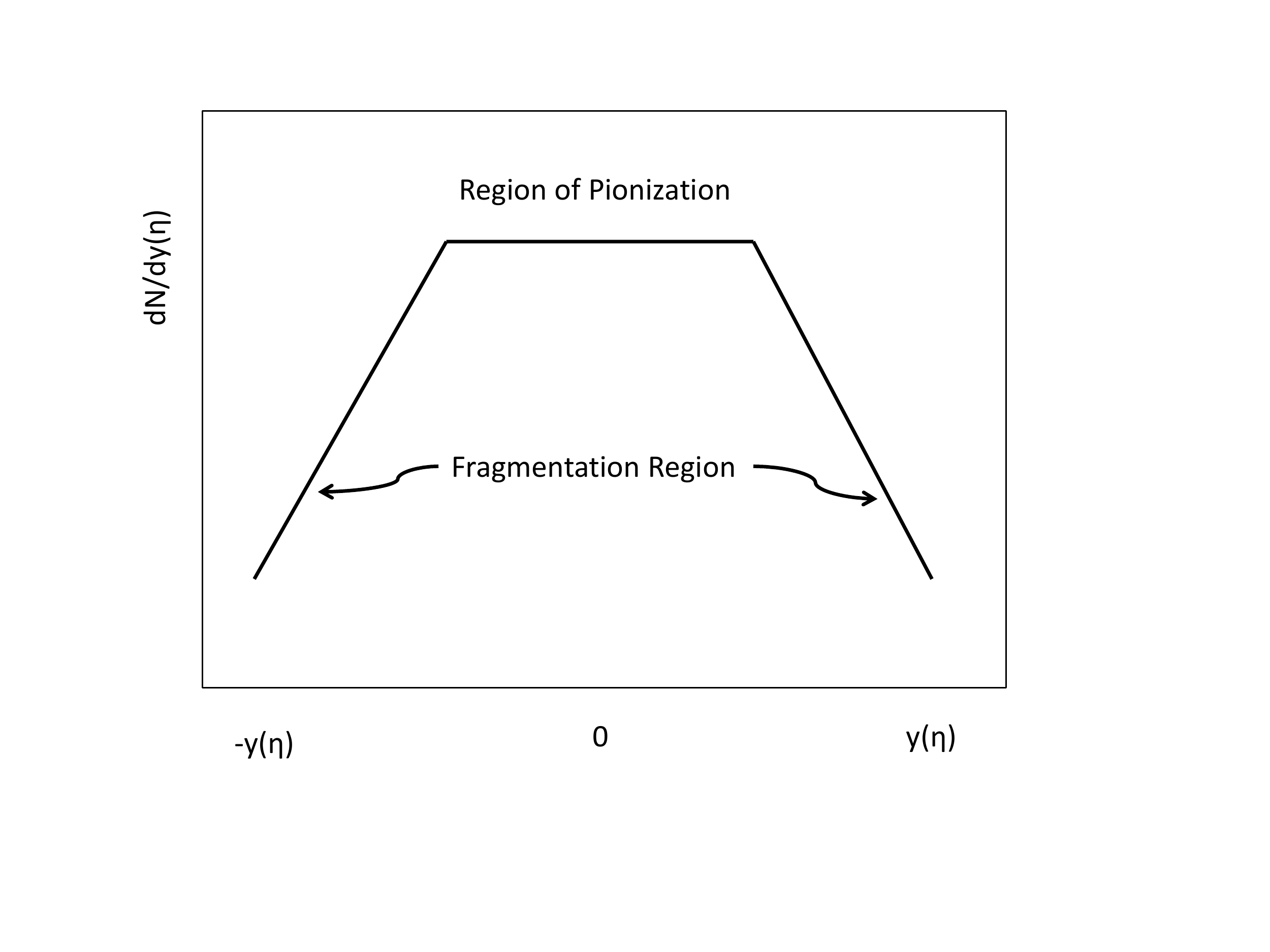}
\caption[]{A schematic of (pseudo)rapidity distribution showing the pionization and fragmentation regions.}
\label{fig1}
\end{figure}

It is expected that a central plateau develops at higher energies, which clearly separates the central rapidity from the fragmentation region. However, as such, there is no separating boundary between the central rapidity and the fragmentation region. The width of the fragmentation region is around 2-units in rapidity~\cite{Nikitin}. The fragmentation region thus, is expected to be well separated from the central region only in very high energies, as the kinematically available rapidity region is much wider than 4-units in rapidity. The particle production in fragmentation region is attributable to the valence quarks participating in hadronization, whereas in central rapidity region, it is dominated by the mid-rapidity gluonic sources at high energies \cite{Wolschin:2013pu,Wolschin:1999jy}. The central rapidity region is called Pionization region ~\cite{Nikitin} and is shown in the Fig.~\ref{fig1}.

There have been several experimental efforts to understand the particle production in both mid and forward rapidities~\cite{Back:2006,UA5,Alver:2009,Adams:2005aa,Adams:2005cy,Abelev:2010,BackR:2005,phobos-prl-91}. As LF is the thrust area of this paper we focus on the particle production in the forward rapidity region. The experimental observation of LF was first reported by the PHOBOS experiment at RHIC with charged particles~\cite{phobos-prl-91}, later STAR experiment also confirmed the hypothesis with inclusive photons in the forward rapidity~\cite{Adams:2005aa}. The Limiting fragmentation was observed by UA5 experiment at CERN for pp and p{$\bar {\rm p}$} collisions from 53 GeV to 900 GeV~\cite{ALICE:2012xs}. However, ALICE experiment at the LHC has reported a violation of LF hypothesis for inclusive photons in pp collisions with limited forward rapidity coverage~\cite{ALICE:2014rma}.  
 
Various theoretical works~\cite{Gelis:2006tb,JalilianMarian:2002wq,bialas,Ruan:2010,Nasim:2011,Bleibel:2016,Stasto:2011zza,Torrieri:2009fv} have reported  the observation of limiting fragmentation phenomenon in heavy-ion collisions. Recently, limiting fragmentation in the era of RHIC and LHC has got a special mention with a new concept called the hypothesis of ``energy-balanced limiting fragmentation" \cite{Sarkisyan:2015gca,Sarkisyan-Grinbaum:2018yld}. In Ref.~\cite{Gelis:2006tb}, it is claimed that the cross-section plays an important role in fragmentation regions. Marian~\cite{JalilianMarian:2002wq} has shown that the LF phenomenon is observed in the differential cross-section per unit pseudorapidity in proton+nucleus collisions at RHIC energies. 
\par
 
Our main aim in this work is to study the phenomena of LF for A+A collisions in view of increasing inelastic particle production cross-section from RHIC to LHC energies. The hypothesis of limiting fragmentation can be tested for both the observables, namely the particle multiplicity density and also the differential cross-section. As LF is least explored in the case of differential cross-section, this work focuses on the later observable with a detailed discussion on multiplicity as well, for a clear comparison of the expected results at the LHC energies. The total hadronic cross-section does not remain constant from lower RHIC energies to the highest LHC energy but is a slowly increasing function of $\sqrt{s}$~\cite{Loizides:2017ack}. The particle production in heavy-ion collisions depends on the hadronic cross-section. Thus, a detailed study of the longitudinal scaling behaviour in terms of cross-section could be a prudent attempt. The longitudinal variables are expected to be sensitive to the available energy and the multiplicity of the produced secondaries. In this context, the study of possible longitudinal scaling of the final state multiplicity as a function of collision energy becomes judicious, in view of increasing inelastic particle production cross-sections at LHC energies. The paper is organised as follows: in Sec.~\ref{LD}, we recapitulate the basics of Landau hydrodynamics and its connection with the limiting fragmentation hypothesis. In Sec.~\ref{LF}, we present the methodology to calculate the differential cross-section per unit pseudorapidity and discuss the results obtained using experimental data and AMPT. Finally, we summarise our findings in Sec.~\ref{CO}.

\section{Landau Hydrodynamics and Limiting Fragmentation Hypothesis}
\label{LD}
The angular distribution of the particles produced in high-energy collisions is described by the famous Landau model with relativistic hydrodynamics given by the conservation of energy momentum tensor, $\partial_{\mu} T^{\mu\nu} = 0$ with a blackbody equation of state, $p= \epsilon/3$, $p$ is the pressure and $\epsilon$ is the energy density ~\cite{Landau:1953gs,Carruthers}. Landau hydrodynamical model assumes complete thermalization of the total energy in the Lorentz contracted volume of the fireball, which makes the initial energy density to grow with collision energy~\cite{Steinberg:2004vy}. The formulation given in~\cite{Steinberg:2004vy} gives rise to the initial entropy of the system, which is produced in the thermalization process of the quanta of the system, to follow a Gaussian distribution in the rapidity space. The width of the rapidity distribution is determined by the Lorentz contraction factor and is related to the speed of sound~\cite{Steinheimer:2012bp}. The multiplicity distribution in the rapidity space, thus becomes \cite{Landau:1953gs,Bjorken:1982qr,Carruthers}
\begin{equation}
\frac{dN}{dy} = \frac{Ks^{1/4}}{\sqrt{2\pi L}} \exp \left(-\frac{y^2}{2L}\right),
\label{landau}
\end{equation}
where $L=\sigma_y^2 = (1/2) \ln(s/m_p^2) = \ln (\gamma)$. Eq.~\ref{landau} can be rewritten as
\begin{equation}
\frac{dN}{dy} = \frac{Ks^{1/4}}{\sqrt{2\pi y_{\rm beam}}} \exp \left(-\frac{y^2}{2y_{\rm beam}}\right).
\end{equation}

The conclusion from Ref.~\cite{Steinberg:2004vy} shows that the hypothesis of limiting fragmentation comes naturally in Landau's model of multiparticle production. Following the LF hypothesis, when the rapidity distribution is seen from one of the projectiles' rest frame, i.e. by transforming  to $y'=y-y_{\rm beam}$, the above expression for rapidity distribution becomes ($dN/dy = dN/dy'$)~\cite{Steinberg:2004vy},
\ba
\frac{dN}{dy'} &=& \frac{Ks^{1/4}}{\sqrt{2\pi y_{\rm beam}}} \exp \left(-\frac{(y'+y_{\rm beam})^2}{2y_{\rm beam}}\right), \nonumber \\
&=& \frac{Ks^{1/4}}{\sqrt{2\pi y_{\rm beam}}} \exp -\left(\frac{y'^2}{2y_{\rm beam}}+y'\right) \exp \left(\frac{-y_{\rm beam}}{2}\right), \nonumber \\
&=& \frac{1}{\sqrt{y_{\rm beam}}} \exp \left(-\frac{y'^2}{2y_{\rm beam}}-y'\right).
\ea
For $y' =0$, the distribution only depends on the Lorentz contraction factor, which is a function of collision energy.
When we make the transformation, $y'=y-y_{\rm beam}$, the fragmentation region shifts by a factor $y_{\rm beam}$, a value which increases with the collision energies, making the region to overlap with each other.

\begin{figure}
\includegraphics[height=18em]{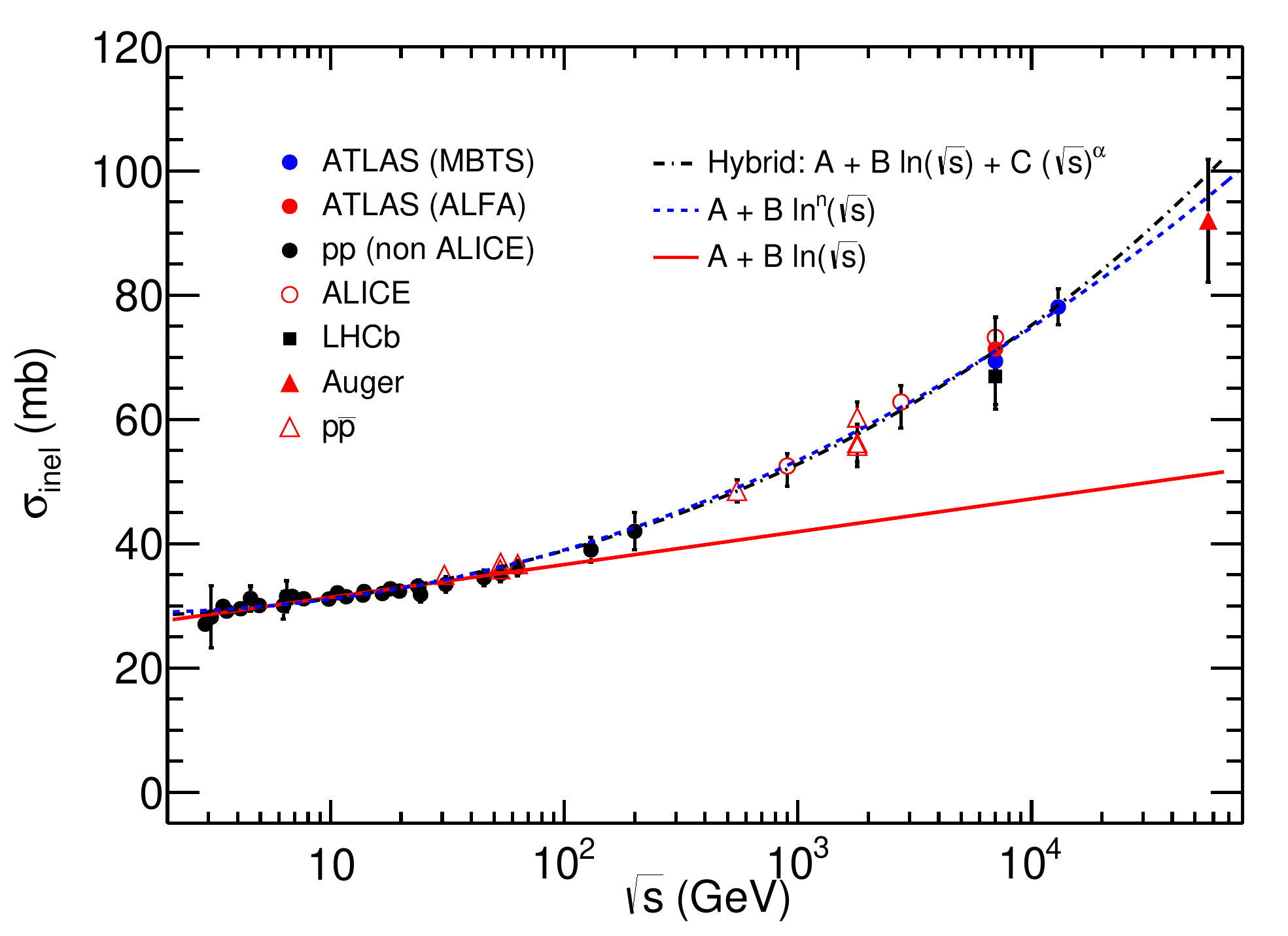}
\caption[]{The inelastic cross-section as a function of $\sqrt{s}$. The symbols are experimental data~\cite{Aad:2011eu,Back:2004je,Abelev:2013qoq,Adam:2015ptt} and the fitted lines are phenomenologically motivated functions.}
\label{sig}
\end{figure}

\section{Limiting Fragmentation at the LHC}
\label{LF}
In this section, we study the limiting fragmentation phenomenon in the pseudorapidity distributions of differential cross-section of charged particles ($d\sigma/d\eta$) for A+A collisions at various center-of-mass energies starting from 19.6 GeV to 5.02 TeV. Due to lack of experimental data of $d\sigma^{AA}/d\eta$, we take the experimentally measured $dN_{\rm ch}^{AA}/d\eta$ at various collision energies. We transform $dN_{\rm ch}^{AA}/d\eta$ into $d\sigma^{AA}/d\eta$ using nucleon-nucleon inelastic cross-sections ($\sigma_{\rm in}$) for different energies applying the method discussed below. A very detailed study is needed to make the connection possible. Recent studies~\cite{JalilianMarian:2002wq} shows that the longitudinal scaling of the differential cross-section per unit pseudorapidity is observed in the experimental data for higher RHIC energies. The rationale behind our work is to bring in the direct center-of-mass energy dependence of $\sigma_{\rm in}$, which has a different low-energy behaviour up to the top RHIC energy in comparison to the LHC energies. This is also observed from the experimentally measured values of $\sigma_{\rm in}$~\cite{Aad:2011eu,Back:2004je,Abelev:2013qoq,Adam:2015ptt}, which are shown in the Fig.~\ref{sig}. In this figure, we show the variation of $\sigma_{\rm in}$ with collision energy. It is clearly seen that there is a very slow rise of $\sigma_{\rm in}$ at lower collision energies up to the top RHIC energy. We have fitted the experimental data with various phenomenologically motivated functions in order to understand the energy-dependent behaviour of $\sigma_{\rm in}$. A logarithmic function, A + B\;ln($\sqrt{s}$), with A and B as free fitting parameters explains the data only up to RHIC energies. This seems to deviate completely after the top RHIC energy. The $\sigma_{\rm in}$ data beyond the top RHIC energy do not follow a logarithmic behaviour. To study the complete energy-dependent behaviour, we have used a hybrid function, A + B ln$(\sqrt{s})$ + C($\sqrt{s})^\alpha$, which combines logarithmic and a power-law to fit the data. Here A, B, C and $\alpha$ are free parameters. This hybrid function explains the data from lower to higher energies. We have also fitted the data with a function A + B\;ln$^{\rm n}(\sqrt{s})$, where A and B are free parameters. A more detailed discussions could be found in Ref.~\cite{Loizides:2017ack}. This seems to describe the data very well. These findings suggest that the logarithmic function alone cannot explain the data for higher energies, while the power of logarithmic function and the hybrid function mentioned above could explain from lower to higher energies shown in the figure. The $\sigma_{\rm in}$ at LHC energies showing a different functional behaviour than the lower energies necessitates a relook into the hypothesis of limiting fragmentation.
\par
 Considering the crude approximation to the physical situation in the framework of  Landau hydrodynamical model of particle production, the relationship between the differential cross-section per unit pseudorapidity ($d\sigma^{\rm pp}/d\eta$) and the pseudorapidity distribution ($dN^{\rm pp}_{\rm ch}/d\eta$) of charged particles for pp collisions is given as \cite{Carruthers:1973ws},

\begin{equation}
\label{sigNch}
\frac{d\sigma^{\rm pp} }{d\eta}= \sigma_{\rm in}\left(\frac{dN^{\rm pp} _{\rm ch}}{d\eta}\right).
\end{equation}
Now, the relation of charged particle pseudorapidity distribution in A+A collisions with the charged particle pseudorapidity distribution in pp collisions using a two-component model, where the contributions from soft and hard processes in the particle production are taken separately, is given as~\cite{Wang:2000bf,Mishra:2015pta},

\begin{equation}
\label{twocom}
\frac{dN^{\rm AA}_{\rm ch}}{d\eta} = \frac{dN^{\rm pp}_{\rm ch}}{d\eta}\left((1-x)\frac{< N_{\rm part}>}{2} + x<N_{\rm coll}> \right).
\end{equation}

Here, $x$ and $(1 - x)$ are the fractions of contribution to the particle production from hard and soft processes, respectively. 
\par
Using Eq.~\ref{twocom} in Eq.~\ref{sigNch}, we get a relation between the differential cross-section per unit pseudorapidity in pp collisions and the charged particle pseudorapidity distribution in heavy-ion collisions as follows: 

\begin{equation}
\label{eqmid}
\frac{d\sigma^{\rm pp}}{d\eta} = \frac{\sigma_{\rm in}\left(\frac{dN^{\rm AA}_{\rm ch}}{d\eta}\right)}{\left((1-x)\frac{< N_{\rm part}>}{2} + x<N_{\rm coll}>  \right)}.
\end{equation}

Now, we proceed towards deriving relationship between differential cross-section per unit pseudorapidity in pp collisions with that in A+A collisions. The distribution of quarks and gluons in a nucleus is different from that in a nucleon with a small effect $(< 10\%)$ of shadowing and EMC effects~\cite{Frankfurt:1991td}. With a crude approximation one can assume that the gluon distribution in a nucleus is just A times that for a proton, where A is the atomic number. The production is expected to increase by a factor of A$^2$ when two nuclei of atomic number A collide in a central way and the pseudorapidity spectrum transforms as~\cite{RChwa},

\begin{equation}
\label{sigma}
\frac{d\sigma^{\rm AA}}{d\eta}= {\rm A}^2\left(\frac{d\sigma^{\rm pp}}{d\eta}\right).
\end{equation}

Using Eqs.~\ref{eqmid} and~\ref{sigma}, we write the differential cross-section per unit pseudorapidity in terms of charged particle pseudorapidity distribution for the heavy-ion collisions as, 

\begin{equation}
\label{sigAA}
\frac{d\sigma^{\rm AA}}{d\eta} = \frac{{\rm A}^2 \sigma_{\rm in}\left(\frac{dN^{\rm AA}_{\rm ch}}{d\eta}\right)}{\left((1-x)\frac{< N_{\rm part}>}{2} + x<N_{\rm coll}>  \right)}.
\end{equation}

\begin{table*}[tp]
 \centering
  \caption{The values of parameters obtained from the fitting of experimental data of $dN_{\rm ch}/d\eta$ with the double Gaussian function given by Eq.~\ref{dGaus}}
  \label{tab:table1}
  \begin{tabular}{c|c|c}
 \hline
Parameters & $\sqrt {s_{\rm NN}}$ = 2.76 TeV          &  $\sqrt {s_{\rm NN}}$ = 5.02 TeV    \\
  \hline
  $A_1$  & 2592.29 $\pm$ 311.56 	& 2102.16 $\pm$ 28.39		\\
  \hline
  $A_2$  & 959.59 $\pm$ 304.26   	& 1817.56 $\pm$ 37.90		\\
  \hline
  $\sigma_1$  & 3.27 $\pm$ 0.13    	& 4.75 $\pm$ 0.01	\\
  \hline
  $\sigma_2$  & 1.67 $\pm$ 0.23   	& 0.61 $\pm$ 0.14		\\	
 \hline
\end{tabular}
\label{t1}
\end{table*}

\begin{figure}
\includegraphics[height=18em]{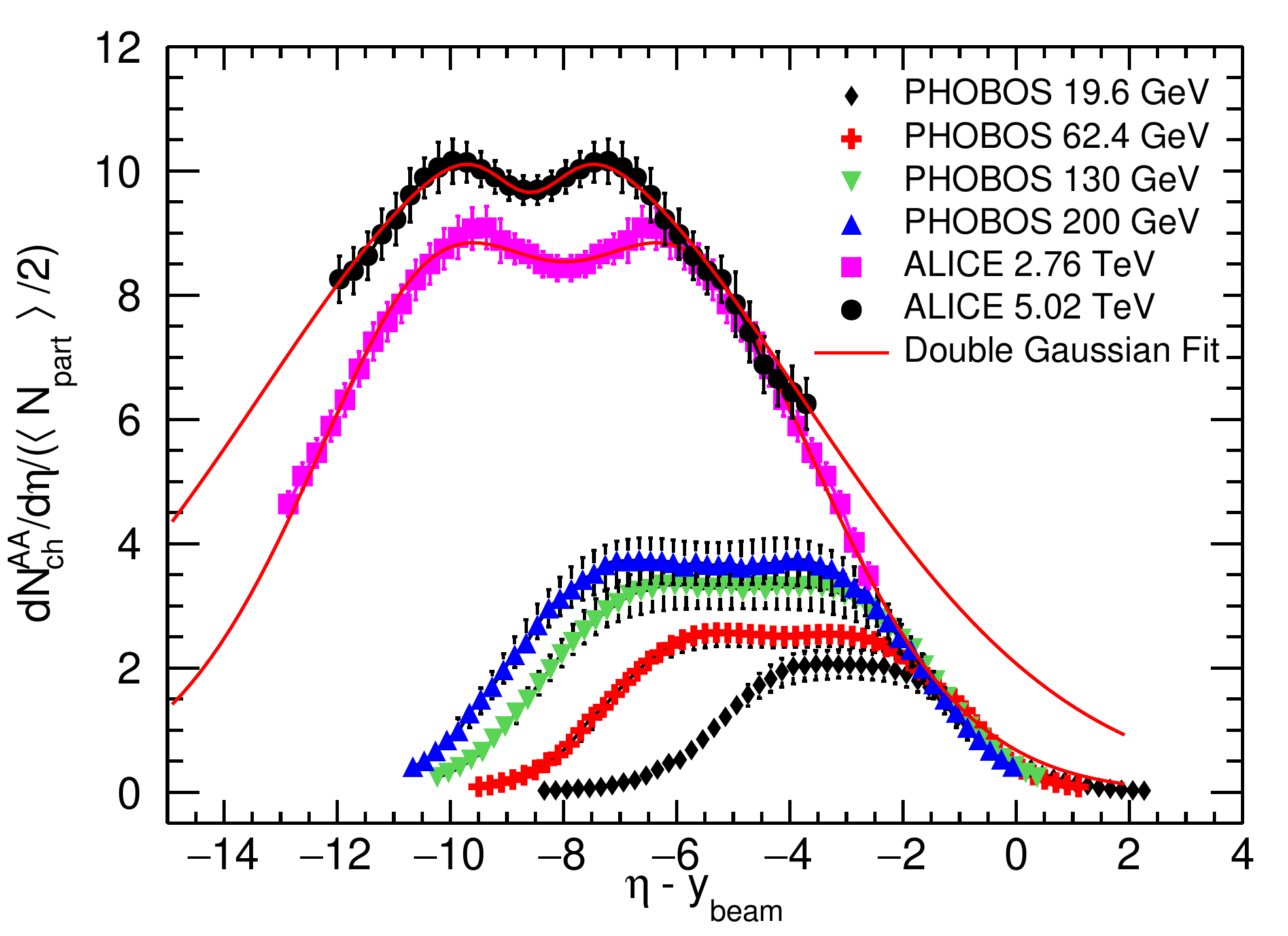}
\caption[]{The number of participant pair normalized pseudorapidity distribution of charged particles ($dN^{\rm AA}_{\rm ch}/d\eta$) in heavy-ion collisions versus $\eta - y_{\rm beam}$ for various energies. The symbols are experimental data~\cite{Back:2004je,Back:2005hs, Abbas:2013bpa, Adam:2016ddh} and the lines are the double Gaussian fits.}
\label{figNpart}
\end{figure}
\par
A large number of experimental data on the charged particle pseudorapidity distribution are available at various center-of-mass energies ranging from RHIC energies like $\sqrt{s_{\rm NN}}$ = 19.6, 62.4, 130 and 200 GeV to LHC energies such as $\sqrt{s_{\rm NN}}$ = 2.76 and 5.02 TeV~\cite{Back:2004je,Back:2005hs, Abbas:2013bpa, Adam:2016ddh}. In a recent paper by the ALICE experiment ~\cite{Abbas:2013bpa}, the limiting fragmentation phenomenon is studied in the pseudorapidity distribution of charged particles at RHIC and LHC energies.   At $\sqrt{s_{\rm NN}}$ = 2.76 TeV, the authors have used a double Gaussian function to extrapolate the data in the fragmentation region and find that the phenomenon of LF is observed at this energy. 

\begin{figure}
\includegraphics[height=18em]{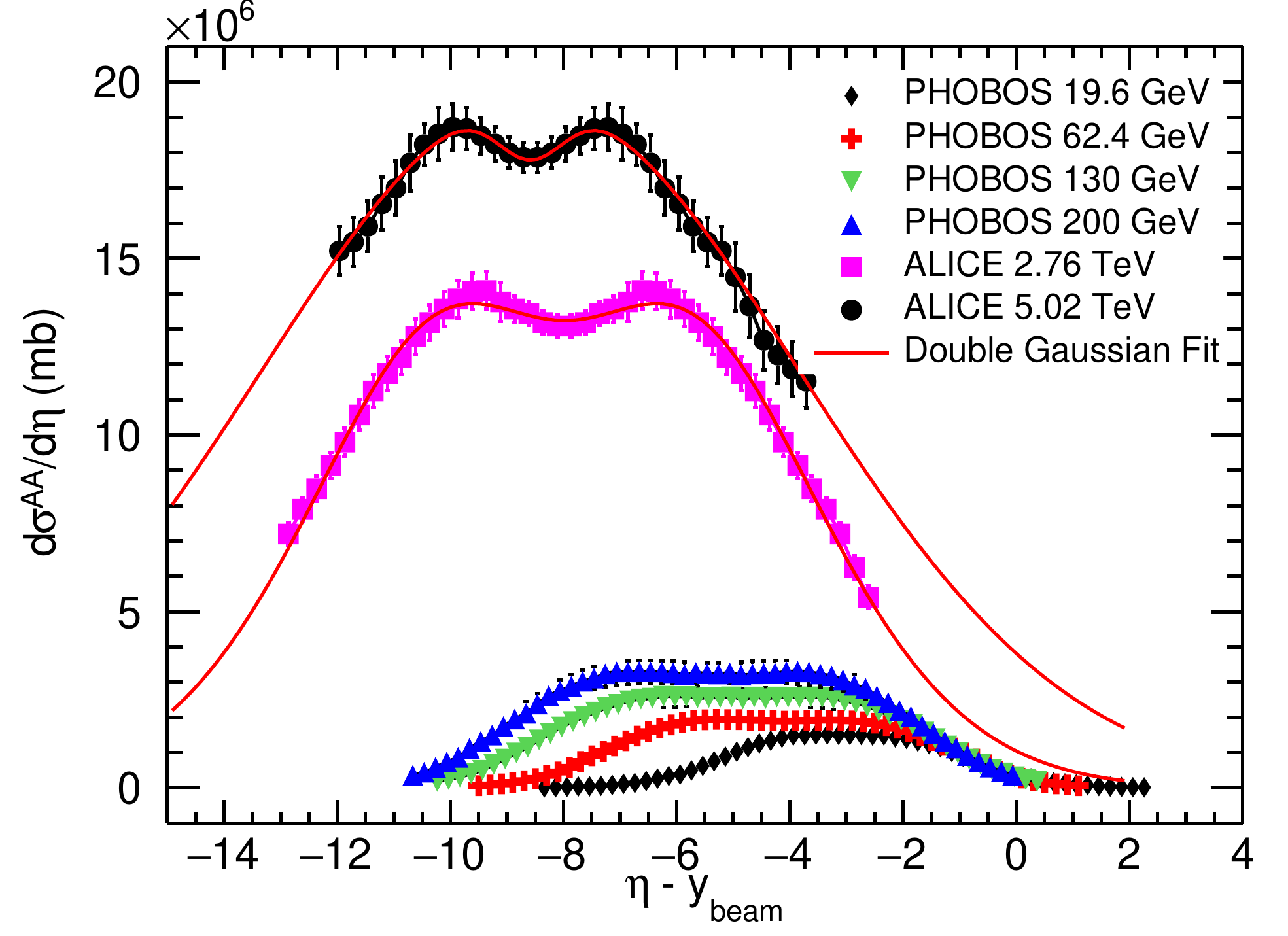}
\caption[]{The differential cross-section per unit pseudorapidity ($d\sigma^{\rm AA}/d\eta$) as a function of $\eta - y_{\rm beam}$ for various collision energies. The symbols are experimental points and the lines are double Gaussian fits.}
\label{figSigma}
\end{figure}

In Fig.~\ref{figNpart}, we have shown $dN^{\rm AA}_{\rm ch}/d\eta/ (<N_{\rm part}>/2)$ as a function of $\eta - y_{\rm beam}$ for various energies from 19.6 GeV to 5.02 TeV. Due to lack of the experimental data in the fragmentation region at LHC energies, we have used double Gaussian function to fit and extrapolate the experimental data in the projectile rapidity region. The double Gaussian function used for fitting is given as follows,

\begin{equation}
\label{dGaus}
f(\eta) = A_1e^{\frac{-\eta^2}{2\sigma_{1}^{2}}} - A_2e^{\frac{-\eta^2}{2\sigma_{2}^{2}}},
\end{equation}
 
where $A_1, A_2$ are the amplitudes and $\sigma_{1}, \sigma_{2}$ are widths of the double Gaussian function. This function describes the experimental data very well at LHC energies within uncertainties ~\cite{Abbas:2013bpa, Adam:2016ddh}. The fitting parameters are given in the table~\ref{t1} for $\sqrt{s_{\rm NN}}$ = 2.76 and 5.02 TeV. We observe that the limiting fragmentation phenomenon seems to be violated at $\sqrt{s_{\rm NN}}$ = 5.02 TeV, while it is observed at energies from $\sqrt{s_{\rm NN}}$ = 19.6 GeV to 2.76 TeV. Despite this, at $\sqrt{s_{\rm NN}}$ = 5.02 TeV, the extrapolation of the charged particle pseudorapidity density scaled with average number of participant does not show a similar behaviour in the fragmentation region as observed at lower energies. The lack of data around the beam rapidity region and the asymmetric values around $\eta = 0$ refrain us to draw any solid conclusion on the behaviour observed at highest LHC energies. It should also be noted here that a Gaussian extrapolation to the fragmentation region is assumption-based and its validity is subjected to a check against the experimental data.

Now, we evaluate $d\sigma^{\rm AA}/d\eta$ using Eq.~\ref{sigAA} for $\sqrt{s_{\rm NN}}$ = 19.6 to 5.02 TeV taking the $x$ parameters from Ref.~\cite{Mishra:2015pta}, which is almost energy independent from RHIC to LHC energies. The inelastic cross-sections for various energies are taken from Ref.~\cite{Aad:2011eu,Back:2004je,Abelev:2013qoq,Adam:2015ptt}. The Monte Carlo Glauber model~\cite{Miller:2007ri} is used to calculate number of participants ($N_{\rm part}$) and number of binary collisions ($N_{\rm coll}$) at different energies. The differential cross-section per unit pseudorapidity for various center-of-mass energies starting from $\sqrt{s_{\rm NN}}$ = 19.6 to 5.02 TeV are shown in Fig.~\ref{figSigma} with respect to $\eta - y_{\rm beam}$. We notice that the limiting fragmentation hypothesis appears to be violated at LHC energies, {\it i.e.} at $\sqrt{s_{\rm NN}}$ = 2.76 and 5.02 TeV. These findings suggest that, it is very important to consider the energy dependent $\sigma_{\rm in}$ in order to study LF phenomenon particularly at LHC energies.

\begin{figure}
\includegraphics[height=18em]{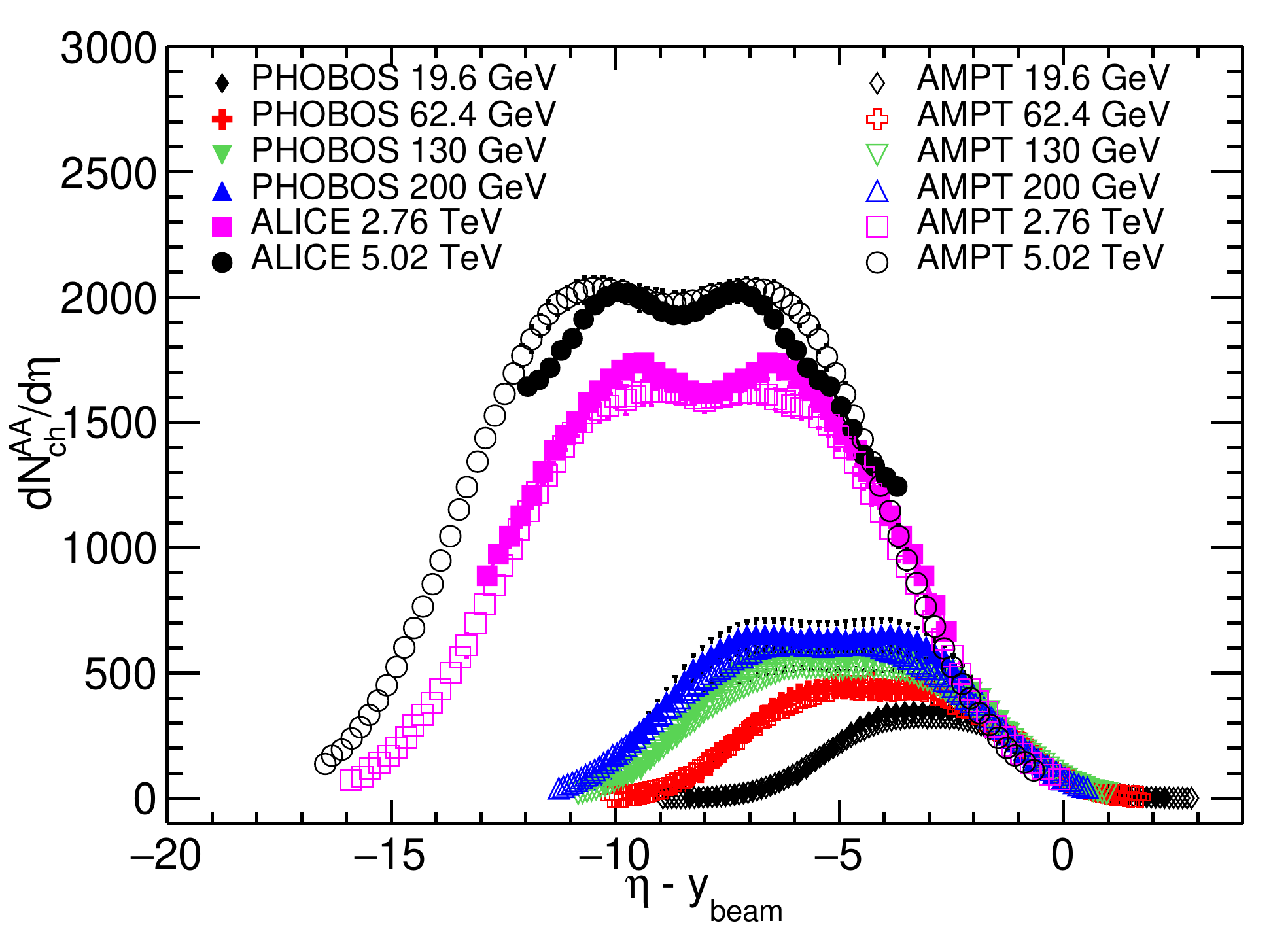}
\caption[]{The comparison of AMPT model predictions with experimental data on $dN^{\rm AA}_{\rm ch}/d\eta$ versus $\eta - y_{\rm beam}$ for various energies.}
\label{figAmpt}
\end{figure}

The experimental data for pseudorapidity distributions of charged particles in the full phase space are not available at the LHC energies. In addition, a double Gaussian extrapolation of $dN_{\rm ch}/d\eta$ to the $y_{\rm beam}$ at a given energy, seem to introduce an artefact in the spectra, which forbids one to look into the hypothesis of limiting fragmentation. To circumvent this problem, we take AMPT model in string melting scenario~\cite{Lin:2004en} as tuned in Ref.~\cite{Basu:2016dmo} for the most central bin 0-6\% and 0-5\% for RHIC and LHC energies, respectively. We have then compared the measured experimental data for pseudorapidity distribution of charged particles~\cite{Back:2004je,Back:2005hs, Abbas:2013bpa, Adam:2016ddh} with the results obtained in AMPT model. The comparison of experimental data with the AMPT model prediction is shown in Fig.~\ref{figAmpt}. AMPT predictions reproduce the mid-rapidity and the fragmentation region very well but cannot reproduce around the peak region ($\eta \sim 0$) at RHIC energies. For LHC energy at $\sqrt{s_{\rm NN}}$ = 2.76 TeV, the AMPT predictions are in good agreement with the experimental data except for the mid-rapidity region, where the predictions slightly underestimate the measured data. Similarly, for $\sqrt{s_{\rm NN}}$ = 5.02 TeV, the predictions from AMPT model slightly overestimate the data measured for 0-5\% centrality bin. In this figure, we see that the phenomenon of longitudinal scaling is observed at RHIC and LHC energies. Theses findings are also described in the Ref.~\cite{Nasim:2011ss}, where various transport models like AMPT and the Ultra-relativistic Quantum Molecular Dynamics (UrQMD) model are used to study this phenomenon. They observed that AMPT (both default and string melting versions) and UrQMD with default version show the longitudinal scaling in pseudorapidity distributions of charged particles at RHIC and LHC energies.

\begin{figure}
\includegraphics[height=18em]{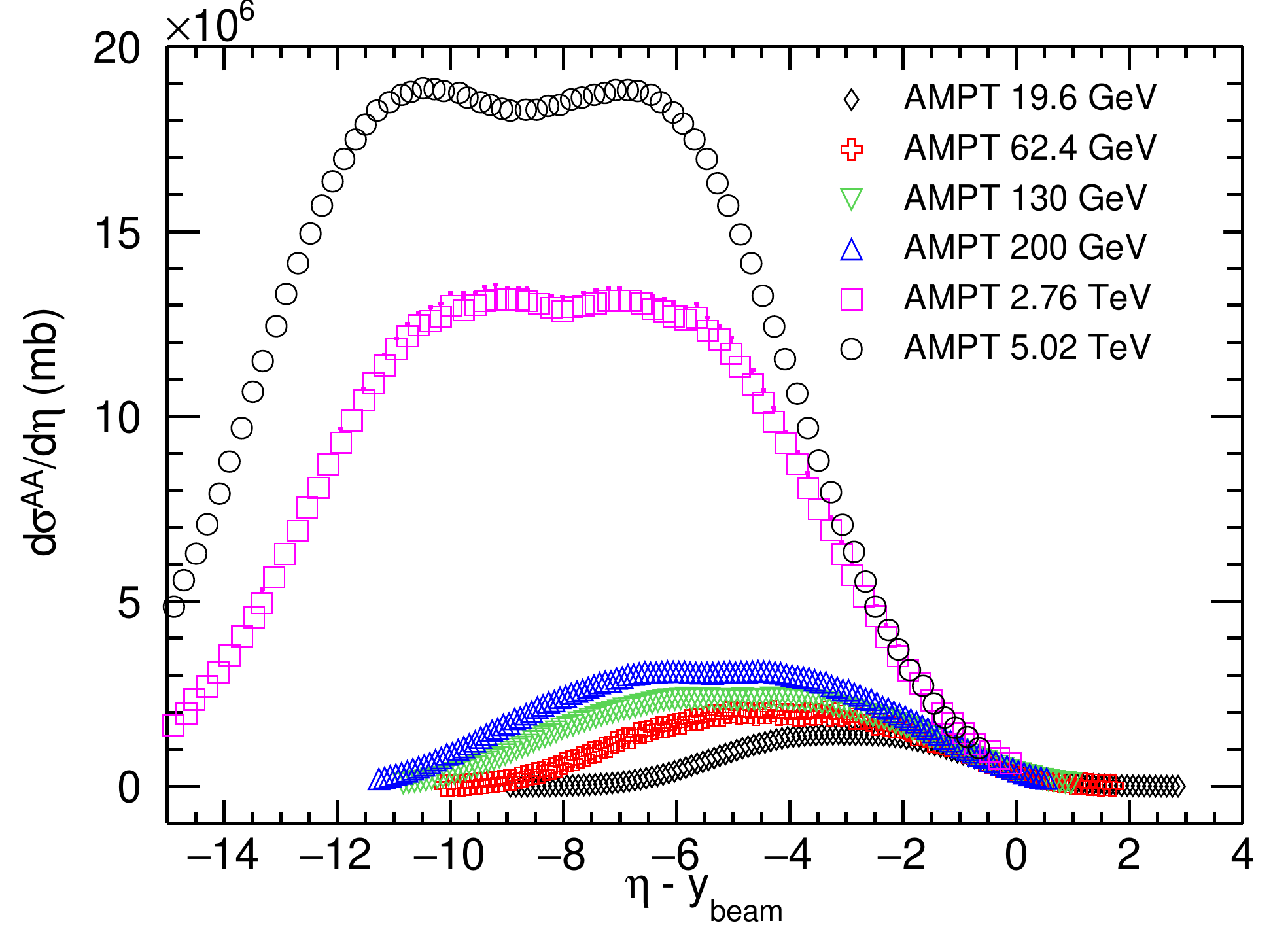}
\caption[]{$d\sigma^{\rm AA}/d\eta$ versus $\eta - y_{\rm beam}$ using AMPT results.}
\label{figAmptCom}
\end{figure}

We convert the AMPT results of $dN^{\rm AA}_{\rm ch}/d\eta$ into $d\sigma^{\rm AA}/d\eta$ using Eq.~\ref{sigAA}. In Fig.~\ref{figAmptCom}, we have shown $d\sigma^{\rm AA}/d\eta $ versus $\eta - y_{\rm beam}$ to see the longitudinal scaling phenomena in the fragmentation region for different energies from 19.6 GeV to 5.02 TeV. Again, we have found a similar observation for the AMPT model as observed in the experimental data i.e. LF is observed up to RHIC energies in $d\sigma^{\rm AA}/d\eta$ and seems to be violated for LHC energies. Theses findings are very important while discussing the longitudinal scaling hypothesis at LHC energies.

\section{Conclusions and Outlook}
\label{CO}
In this work, we have revisited the phenomenon of limiting fragmentation in the pseudorapidity distributions of differential cross-sections of the charged particles using the energy dependent inelastic cross-section. The findings of this analysis are: 

\begin{itemize}

\item We have observed the limiting fragmentation phenomenon in the experimental data of $dN^{\rm AA}_{\rm ch}/d\eta$ from $\sqrt{s_{\rm NN}}$ = 19.6 GeV to 2.76 TeV and it is violated at $\sqrt{s_{\rm NN}}$ = 5.02 TeV. Here, the double Gaussian function is used to extrapolate the experimental data in the fragmentation region. However, on the basis of extrapolation method, one can not infer any exact physics conclusions.

 \item We have transformed experimental data of $dN^{\rm AA}_{\rm ch}/d\eta$ to $d\sigma^{\rm AA}/d\eta$ for various energies from $\sqrt{s_{\rm NN}}$ = 19.6 GeV to 5.02 TeV and see the distributions in the rest frame of one of the nuclei. We have found that the LF hypothesis seems to be violated at both the energies i.e. at $\sqrt{s_{\rm NN}}$ = 2.76 and 5.02 TeV, when one considers the energy dependent inelastic cross-section. 
 
 \item We have also studied the phenomenon of longitudinal scaling using AMPT model and employing the same procedure as used for the experimental data. Our studies suggest that, AMPT seems to show a possible violation of limiting fragmentation phenomenon for $d\sigma^{\rm AA}/d\eta$ at LHC energies.  
 
\item {The hypothesis of LF comes as a natural outcome when the particle production follows the Landau hydrodynamics, with a Gaussian pseudorapidity profile.} 

\item{LF works fine, when the hadronic cross-section is assumed to be almost independent of energy, which is not the case and hence it is expected to be violated at higher energies. We find that the limiting fragmentation appears to be violated at LHC energies while using the energy dependent cross-section.}

\item{The thermal model with Landau extrapolation to LHC for charged particles, predicts a violation of LF at LHC \cite{Cleymans:2007jj}. What about photons in this framework? It has been observed that for pions in thermal model with longitudinal flow, the LF is violated at the LHC energies \cite{Tiwari:2016ovw}. What about photons with a longitudinal flow? These need further investigations.}

\item{It is expected that at higher energies, Landau hydrodynamics should fail and we should expect Bjorken boost invariant hydrodynamics to work out, with the observation of a mid-rapidity plateau.  If LF is a natural outcome of Landau model, then LF should be violated at LHC for two reasons: i) failure to see a Gaussian pseudorapidity distribution and ii) cross-sections vary substantially towards higher collision energies}.

\item{At lower collision energies, baryon stopping at the mid-rapidity is expected and the $dN_{\rm ch}/d\eta (y)$ is expected to follow a Gaussian-like behaviour, which could be described by the particle production in Landau hydrodynamic model. Hence, at these energies, the observation of a limiting fragmentation hypothesis in particle production is expected. But at higher energies, where Landau hydrodynamics fails due to the absence of Gaussian rapidity distribution, LF is found to be violated.}

\item{Going from the top RHIC energy to the LHC energies, there is an order of magnitude increase in the collision energy. Considering at least two units of (pseudo)rapidity overlap for the LF to be valid, the observed $y_{\rm beam}$ at $\sqrt{s_{\rm NN}}$ = 200 GeV and 5.02 TeV makes hardly any overlap in (pseudo)rapidity. While looking into the possible observation of limiting fragmentation, one looks at spectral overlap in the fragmentation region, which may not be expected as mentioned. Hence, RHIC can't be combined with LHC while looking for the hypothesis of Limiting Fragmentation.}

\item Theoretical models are mostly assumption dependent. In order to validate a model, one needs to confront a model to experimental data. We need forward charged particle and photon detectors at the LHC energies in order to validate the LF hypothesis. In the absence of this, extrapolation of any theoretical findings from mid-rapidity to extreme forward rapidity would be a speculation sometimes or a mere coincidence, as the physics of particle production is highly rapidity dependent. In view of this, in the present work we have taken the inelastic cross-section with the collision geometry to study the LF hypothesis. This is the novelty of the present work.
\end{itemize}

\
 
\section*{ACKNOWLEDGEMENTS}
RS acknowledges stimulating discussions with Edward Sarkisyan-Grinbaum. Useful helps from Aditya Nath Mishra while preparing the manuscript is highly appreciated. The authors acknowledge the financial supports from ALICE Project No. SR/MF/PS-01/2014-IITI(G) of Department of Science \& Technology, Government of India. This research work used resources of the LHC grid computing centre at Variable Energy Cyclotron Center, Kolkata.

\end{document}